\documentclass[12pt]{article}

\usepackage[colorlinks=true, urlcolor=blue, linkcolor=blue, citecolor=blue]{hyperref}
\usepackage{dsfont}
\usepackage{amssymb}
\usepackage{amsmath}
\usepackage{mathtools}
\usepackage{graphicx}
\usepackage{enumerate}
\usepackage[english]{babel}
\usepackage{booktabs}
\usepackage{pdflscape}
\usepackage{natbib} 
\usepackage{microtype} 
\usepackage{subcaption}
\usepackage{dblfloatfix}
\usepackage[margin=1in]{geometry}

\usepackage{array}

\newcommand{\R}{\mbox{$\mathbb{R}$}}

\newcommand{\N}{\mbox{$\mathbb{N}$}}

\newcommand{\whsigma}{\widehat{\sigma}}
\newcommand{\bqan}{\begin{eqnarray}}
\newcommand{\eqan}{\end{eqnarray}}
\newcolumntype{L}[1]{>{\raggedright\let\newline\\\arraybackslash\hspace{0pt}}m{#1}}

\newcommand\BibTeX{{\rmfamily B\kern-.05em \textsc{i\kern-.025em b}\kern-.08em
		T\kern-.1667em\lower.7ex\hbox{E}\kern-.125emX}}
\makeatletter
\let\@fnsymbol\@arabic
\makeatother

\begin{document}
		\definecolor{ao(english)}{rgb}{0.0, 0.5, 0.0}
		\definecolor{bronze}{rgb}{0.8, 0.5, 0.2}
		\definecolor{byzantine}{rgb}{0.74, 0.2, 0.64}

	\title{\textbf{A cautionary tale on using imputation methods for inference in matched pairs design.}}

	\author{Burim Ramosaj$^*$\thanks{Ulm University, Institute of Statistics, Ulm, Germany.\newline \text{ \hspace{0.42cm}}  *Corresponding author. E-mail: burim.ramosaj@uni-ulm.de}  \text{\hspace{0.0001cm}},  Lubna Amro\footnotemark[1]  and Markus Pauly\footnotemark[1]}
	
	\maketitle

	\begin{abstract}
		\textbf{Motivation:} Imputation procedures in biomedical fields have turned into statistical practice, since further analyses can be conducted ignoring the former presence of missing values. In particular, non-parametric imputation schemes like the random forest or a combination with the stochastic gradient boosting have shown favorable imputation performance compared to the more traditionally used MICE procedure. However, their effect on valid statistical inference has not been analyzed so far. This paper closes this gap by investigating their validity  for inferring mean differences in incompletely observed pairs while opposing them to a recent approach that only works with the given observations at hand.\\
		\textbf{Results:} Our findings indicate that machine learning schemes for (multiply) imputing missing values may  inflate type-I-error or result in comparably low power in small to moderate matched pairs, even after modifying the test statistics using Rubin's multiple imputation rule. In addition to an extensive simulation study, an illustrative data example from a breast cancer gene study has been considered.\\
		\textbf{Availability:} The corresponding \texttt{R}-code can be accessed through the authors and the gene expression data can be downloaded at www.gdac.broadinstitute.org\\
	\end{abstract}
	

	\section{Introduction}
	\label{sec1}
	
Repeated measure designs are commonly present in medical and scientific research. The simplest design captures the comparison of two different treatments or time points over repeatedly observed subjects. For inferring the predominance of a treatment the paired t-test is commonly applied. It is finitely exact under normality and asymptotically exact, if the latter assumption is dropped. However, first limitations of the paired t-test arise when missing values are present. This is frequently the case in biomedical studies, e.g. due to dropout or insufficient probes. In addition, rare diseases often lead to small sample sizes in matched pairs. Hence, conducting complete-case analyses through the paired t-test while neglecting all information from the incomplete observations might lead to improper inferences, especially when missing values are not occurring completely random resulting into selection bias or when sample sizes are small \citep{schafer1999multiple}. Therefore, scientific research has been going on involving missing value mechanisms in the test statistic.  One simple approach is to impute missing values singly and conduct statistical tests as if there has not been any missing values so far. However, this procedure fails to account for the uncertainty involved in the imputation technique and statistical inference can be heavily distorted \citep{schafer1999multiple,rubin2004multiple, sterne2009multiple}.  Multiply imputing missing values overcome this issue and the combination of them into a potentially asymptotic exact test statistic has been derived by \cite{rubin2004multiple}. Various algorithms have been developed to impute missing values. For example, the \texttt{mice} package in \textsf{R} is able to impute missing values multiple times by assuming a parametric model for the data generating process and iteratively draw through fully conditional specification missing values from a predictive posterior distribution. Recently, non-parametric regression and classification methods such as the random forest have been proposed for the imputation of missing values \citep{stekhoven2011missforest}. In extensive simulation studies \citep{stekhoven2011missforest, waljee2013comparison, ramosaj2017wins}, the \texttt{missForest} approach accurately imputes missing values and outperforms existing procedures in simulations such as MICE in terms of normalizing root mean squared or miss-classification error. 
Thus, we would expect that inference drawn from the imputed data might lead to reliable conclusions. 
However, as recently pointed out by \cite{dunson2018statistics} for machine learning techniques in general, 
{\it 'there is a huge danger in doing so due to the lack of uncertainty quantification'}. In the present paper, we investigate this issue by checking statistical validity for drawing inference after multiple imputation. 

Differing to imputation, research has also been going on in developing statistical testing procedures that only take into account the observed information \citep{lin1974difference,ekbohm1976oncomparing,bhoj1978testing,looney2003method,kim2005statistical, samawi2014notes, maritz1995permutation, yu2012}. In particular, statistical permutation tests were provided recently by \cite{amro2017permuting} and \cite{amro2018multiplication} that require no parametric assumptions and possess nice small and large sample properties while only using all observed information. Below we will compare them to the previously mentioned inference strategies based on imputation techniques.

Similar studies on classical imputation techniques have shown that inference based on classical multiple imputation techniques and incomplete data-analysis can differ, arising the question under which condition multiple imputation lead to correct statistical inference \citep{fay1992inferences, meng1994multiple, sterne2009multiple,peto2007doubts}. \cite{meng1994multiple} for example, provided a general framework under which 
an imputation method 
leads to correct inference. He coined the term congeniality for assessing differences in imputation and analysis model under a Bayesian scheme. Under specific conditions defined in his work such as \textit{second-moment properness,  self-efficiency} and \textit{information regularity}, multiple imputation can lead to even more efficient estimates than solely based on incomplete data. However, if the imputation model is not correctly specified, inference is generally not valid. Therefore, the main task of the imputer is to deliver a predictive model that is generalizable, covering potential sub-models in the subsequent analysis while correctly reflecting the uncertainty by the imputation \citep{meng1994multiple}. Due to the flexibility and predictive accuracy of the Random Forest model without making assumptions on the data generating process, this method might be regarded as a potential candidate model for imputing missing values according to \cite{meng1994multiple}. However, less attention has been paid to the interesting question, whether a Random Forest based imputation scheme leads to correct statistical inferences  in e.g. matched pairs design. Therefore, the aim of the current paper is trilateral. First, we aim to  i)  enlighten the mystery of correct statistical inference in matched pairs design when advanced Machine Learning techniques are used for multiply imputing missing values, ii) to check whether imputation accuracy measures such as root mean squared errors are capable of reflecting correct statistical inferences and iii) recommend a valid statistical testing procedure in matched pairs under ignorable missing mechanisms such as missing completely at random. 

After stating the statistical model and procedures of interest in Section $\ref{sec2}$, we describe multiple imputation techniques in Section $\ref{sec3}$. Section $\ref{sec4}$ explains the application of the weighted permutation procedure. In an extensive simulation study (Section $\ref{sec5}$) under various sample sizes and covariance structures, we then compute the type-I-error rate and the power of the paired t-test under different imputation schemes and oppose it to a recently proposed permutation test. The results are discussed in Sections $\ref{sec5_1}-\ref{sec5_3}$ and the analysis of a breast cancer gene study exemplifies potentially diverse findings in Section $\ref{sec6}$.

	\section{Statistical Model and Hypotheses}
	\label{sec2}
	We consider the general matched pairs design given by a sample $\mathcal{D}_n := \{ \mathbf{X}_1, \dots , \mathbf{X}_n \}$, where $\mathbf{X}_j \in \R^2$ are i.i.d. random vectors on some probability space $(\Omega, \mathcal{F}, \mathbb{P})$ with mean vector $\mathbb{E}[\mathbf{X}_1]  = \boldsymbol{\mu} = [\mu_1, \mu_2]^\top \in \R^2 $ and an arbitrary covariance matrix  
	$\boldsymbol{\Sigma} > 0$. Let the vectors $\mathbf{R}_j = [R_{1j}, R_{2j}]^\top$ 
	indicate whether a certain component $X_{ij}$ is observed $(R_{ij} = 1)$ or missing $(R_{ij} = 0)$ on subject $j = 1, \dots, n, $ $i = 1,2$. Denoting component-wise multiplication of vectors by $*$, then we only observe $\mathbf{X}^{(o)} :=  \{  \mathbf{X}_j * \mathbf{R}_j \}_{j = 1}^n$. Hence our framework has the following form, where $---$ stands as  a placeholder for $R_{ij} = 0$: 
	
	\begin{equation}
	\begin{gathered}
	\label{ModelMis}
	\underbrace{ \begin{bmatrix}
		X_{11}^{(c)} \\
		X_{21}^{(c)}
		\end{bmatrix},  \dots ,\begin{bmatrix}
		X_{1n_1}^{(c)} \\
		X_{2n_{1}}^{(c)} 
		\end{bmatrix}}_{ \mathbf{X}^{(c)} }, \\
	\underbrace{
		\begin{bmatrix}
		X_{11}^{(i)}\\
		\text{ }_{---}
		\end{bmatrix}, \dots , \begin{bmatrix}
		X_{1n_{2}}^{(i)} \\
		\text{ }_{---}
		\end{bmatrix} 
		\begin{bmatrix} 
		\text{ }^{---}\\
		X_{21}^{(i)} 
		\end{bmatrix}, \dots , \begin{bmatrix}
		\text{ }^{---} \\
		X_{2n_{3}}^{(i)} 
		\end{bmatrix}.}_{ \mathbf{X}^{(i)} }
	\end{gathered}
	\end{equation}
	
	Here, $\mathbf{X}^{(c)}$ are the complete pairs and $\mathbf{X}^{(i)}$ partially observed pairs. Rubin defines the missing mechanism through a parametric distributional model on $\mathbf{R} = \{ \mathbf{R}_j \}_{j = 1}^n$ and classifies their presence through Missing Completely at Random (MCAR), Missing at Random (MAR) and Missing not at Random (MNAR) schemes \citep{rubin2004multiple}. In our work, we restrict our attention to missing mechanisms being MCAR. 
	To fix some notation, let $I_{n_1}$ denote the index set of the $n_1 = \lvert I_{n_1} \rvert$ complete pairs, i.e. $\mathbf{R}_{j} = [1,1]^\top$ for all $j \in I_{n_1}$. Similarly, $I_{n_2}$ resp. $I_{n_3}$ denote the index sets 
	of observations with missing second $(\mathbf{R}_j = [1,0]^\top$ for $j \in  I_{n_2} )$ resp. first  $(\mathbf{R}_j = [0,1]^\top$ for $j \in I_{n_3} )$ component and set $\lvert I_{n_2} \rvert =n_2$, $\lvert I_{n_3} \rvert =n_3$, $n = n_1 + n_2 + n_3$. 
	In this set-up (\ref{ModelMis}), we are now interested in testing the null hypothesis $H_0:$ $\mu_1 = \mu_2$ of equal means against the two-sided alternative $H_1:$  $\mu_1 \neq \mu_2$. Tests for the one-sided analogue can be achieved by transforming the corresponding two-sided $p$-value to $\max \{ p/2, 1-p/2 \}$.

	\section{Imputing Incomplete Data}
\label{sec3}
A usual habit is to impute missing values in the sample $\mathcal{D}_n$ and conduct further statistical analyzes on the imputed data set $\mathcal{D}_n^{imp}$. Various imputation schemes have been analyzed with respect to their imputation accuracy. Recently, promising results were obtained by a random forest based imputation scheme implemented as \texttt{missForest} in \textsf{R} \citep{stekhoven2011missforest}. It is based on a non-parametric regression model combining several weak learners to construct strongly predictive models while including random subsampling in feature selection and bootstrapping. In several competitive simulation studies \citep{stekhoven2011missforest, waljee2013comparison, ramosaj2017wins}, $\texttt{missForest}$ outperformed different imputation techniques as, e.g., $k$-NN based techniques or the usually applied MICE algorithm. The latter is based on an iterative algorithm generating Markov-Chains for potential imputation values \citep{buuren2010mice}. However, imputation retrieves the risk that imputer and data analyst could share completely different viewpoints for subsequent analyses. While the imputer is interested in using a method that is generalizable and thus minimizes normalized root mean squared error (NRMSE) or proportion of false classification (PFC), the interest of a statistical data analyst lies more on uncertainty quantification and valid inference. From a biostatistical viewpoint, it is unclear whether a method with comparable low NRMSE results in a test-procedure that reasonably controls type-I-error under the null-hypothesis. For this task, an ideal  imputation scheme should be based on preserving the distributional property of the data generating process. That is,  $[X_{1j}^{(c)}, X_{2j}^{(c)}]^\top = \mathbf{X}_{j}^{(c)}  \stackrel{d}{=} \mathbf{X}_j^{imp} = [X_{1j}^{imp}, X_{2j}^{imp}]^\top$ should hold for $j \in I_{NA} :=  I_{n_2} \cup I_{n_3} $, where $\mathbf{X}_j^{imp}$ is the imputed value of the $j$-th observation. In this case, a natural test statistic for $H_0$ under a distribution-preserving imputation scheme would be the paired t-test statistic given by 

\begin{align}
\label{PairedApprox}
T^{comp} &:=  \sqrt{n}\frac{\frac{1}{n}  \mathbf{1}_n^\top \mathbf{d}_n - \mathbf{c}^\top \boldsymbol{\mu}}{ \sqrt{ \frac{1}{n-1} \mathbf{d}_n^\top \mathbf{P}_n \mathbf{d}_n  }  } \stackrel{H_0}{\approx} t_{n-1}.
\end{align}
Here, $\mathbf{1}_n^\top = [1, \dots, 1]^\top$ denotes the $n$-dimensional vector of $1$s, $\mathbf{c}^\top = [-1,1]$ the difference contrasts, $d_l^{(c)} = \mathbf{c}^\top \mathbf{X}_l^{(c)}$, $l \in \{1, \dots, n_1 \}$ and $d_j^{(i)} = \mathbf{c}^\top \mathbf{X}_j^{(imp)}$, $j \in I_{NA}$ the mean differences of the first and second components of the complete and imputed pairs, respectively, which are pooled in the vector $\mathbf{d}_n^\top = [d_1^{(c)}, \dots, d_{n_1}^{(c)}, d_{1}^{(i)}, \dots, d_{n_2}^{(i)}, d_{n_2 +1}^{(i)}, \dots, d_{n_2+n_3}^{(i)}]^\top \in \R^n$. Moreover, building the quadratic form in the matrix $\mathbf{P}_n = \mathbf{I}_n - 1/n \mathbf{1}_n\mathbf{1}_n^\top$ leads to the usual empirical variance estimator in the denominator. 
Althought the stated $t$-null-distribution in \eqref{PairedApprox} is only valid for independent bivariate normal vectors $\mathbf{X}_1, \dots, \mathbf{X}_n$, the paired-t-test is asymptotically exact even under non-normal observations due to the central limit theorem and the law of large numbers.

\subsection{MICE}

Multiple Imputation by Chained Equations, often referred as fully conditional specification, uses a set of conditional parametric models for the imputation of missing values \citep{hughes2014joint}. Imputation is then conducted by introducing a regression model for every variable with a suitable prior and draws are made iteratively form the corresponding predictive posterior of the regression model. That is, for every variable $j \in \{1,2\}$ a separate model with parameter $\theta_j$ is assumed and simulation is conducted by the following iteration steps for $t = 1, 2, \dots, T$, $T \in \N$
\begin{equation*}
\begin{gathered}
\theta_j^{(t)} \sim \mathbb{P}(\theta_1) \mathbb{P}( X_{j1}^{(c)}, \dots X_{jn_1}^{(c)}, X_{j1}^{(i)}, \dots, X_{jn_{j+1}}^{(i)}  | \theta_j^{(t-1)} ) \\
X_j^{*(t)} \sim \mathbb{P}(X_{\pi(j)1}^{(i)}, \dots, X_{\pi(j)n_2}^{(i)} | X_{11}^{(c)}, \dots,  X_{1n_1}^{(c)},\\ 
\hspace{5cm}X_{j1}^{(i)}, \dots , X_{jn_{j+1}}^{(i)}  , \theta_j^{(t)} ),
\end{gathered}
\end{equation*}
where $\pi(1) = 2$ and $\pi(2) = 1$. The final imputation is then the $T$-th draw, i.e. $X_j^{*(T)}$. This allows for a higher flexibility in contrast to joint modeling, where the specification of a parametric joint model  $\mathbb{P}(\mathbf{X}^{(c)}, \mathbf{X}^{(i)} | \theta)$ with joint parameter $\theta$ can be challenging under different variable types. Under the ignorability assumption, imputations in joint modeling using the Bayesian framework are then draws from the posterior predictive distribution of the missing data given the observed observations $\mathbb{P}( X_{21}^{(i)}, \dots, X_{2n_2}^{(i)},, X_{11}^{(i)}, \dots, X_{1n_3}^{(i)}| \mathbf{X}^{(c)}, X_{11}^{(i)}, \dots, X_{1n_2}^{(i)}, X_{21}^{(i)},$ \\$ \dots, X_{2n_3}^{(i)})$ \citep{schafer1997analysis}. In case of finite samples sizes, compatibility and a non-informative margins condition, both methods, MICE and joint modeling, draw from the same predictive distribution of $\mathbf{X}^{(i)}$. This is especially the case if data results from a multivariate normal distribution \citep{hughes2014joint} such that statistical inference should be - at least asymptotically - valid in our case.  \\
Depending on the conditional distribution of $X_{1i} | X_{2i}$ resp. $X_{2i} | X_{1i}$, $i = 1, \dots, n$, different algorithms have been proposed for MICE. In case of normally distributed conditionals, a linear model is assumed and missing values are drawn from it using a multi-stage wild-bootstrap approach \citep{van2006fully}. In the sequel, this approach will be named \textbf{NORM} and is implemented in the \textsf{R}-package \texttt{mice}. Predictive Mean Matching (\textbf{PMM}), is more flexible towards deviations from the normality assumption including nonlinear association and heteroscedasticity \citep{schenker1996partially, morris2014tuning, vink2014predictive, yu2007evaluation}. In contrast to NORM, PMM imputes values by randomly drawing from a set of possible donors, that are determined based on the euclidean distance between observed items and predicted values of missing variables under the same linear model as in NORM \citep{van2006fully}. This way, the regression model is only used for finding appropriate matches between observed and missing cases. 

\subsection{Random Forest Imputation}
Random Forest is a specific class of CART algorithms combining bagging and feature sub-spacing at each terminal node during the tree construction process. Due to its capability of detecting collinearity effects and accommodating nonlinear relations and interactions \citep{shah2014comparison}, the Random Forest has become popular for its usage in missing value estimation \citep{burgette2010multiple, stekhoven2011missforest, doove2014, shah2014comparison}. In several simulation studies and medical data records, the implemented routine \texttt{missForest} in \textsf{R} has been used as a favorite choice for imputation \citep{stekhoven2011missforest, waljee2013comparison, ramosaj2017wins}. Extending the usage of \texttt{missForest} and Random-Forest based imputation schemes to multiple imputations that account for uncertainty, some attempts have been made to incorporate these methods into the MICE framework \citep{doove2014, shah2014comparison}. However, it seems that none of these methods simultaneously considered the validity of statistical inference and imputation accuracy, focusing for example on either high imputation accuracy \citep{stekhoven2011missforest} or the estimation of unbiased regression coefficients in two-way interaction regression models \citep{doove2014}. In this paper, we aim to fill this gap by considering Random Forest based imputation methods within the framework of

\begin{enumerate}[(i)]
	\item Multiple Imputation by repeatedly applying \texttt{missForest} (\textbf{RF MI}) 
	\item Multiple imputation using Random Forest within the MICE framework (\textbf{RF MICE}) \citep{doove2014}
\end{enumerate}

while comparing the corresponding type-I-error and power of the paired t-test and the NRMSE in matched pairs. There are a couple of differences between RF MI and RF MICE, focusing either on more stable prediction of missing values or on extra variability due to imputation. RF MI initially imputes mean resp. mode values for missing values originating from continuous resp. categorical outcomes. RF MICE, however, randomly selects one candidate from the observed part for imputation. In contrast to RF MI, where the prediction of missing values is based on averaging donors in each leaf, RF MICE randomly selects candidates among the donors. In addition, the iteration process in RF MICE is repeated for a fixed number of times, whereas in RF MI, iteration is stopped when a stopping criterion measuring the imputation accuracy is reached. 

\subsection{Rubin's Rule}
In the matched pairs design with imputed missing values, the quantity of interest in the null hypothesis is the difference in expectations $\mathbf{c}^\top \boldsymbol{\mu} = \mu_1 - \mu_2$, where a natural unbiased estimate under a distribution-preserving imputation scheme is given by $\hat{Q}_n :=  1/ n \left( \sum_{j = 1}^{n_1} \mathbf{c}^T\mathbf{X}_j^{(c)} + \sum_{j \in I_{NA}} \mathbf{c}^\top \mathbf{X}_j^{imp}  \right)$. Rubin proposed to impute multiple times, in oder to reflect the uncertainty of the imputation scheme in later inference. Hence, if an imputation scheme estimates missing values $m > 0$ times, leading to $\hat{Q}_n^{(1)}, \dots, \hat{Q}_n^{(m)}$ estimates of mean value differences and $\hat{U}_n^{(1)}, \dots \hat{U}_n^{(m)}$ of empirical variances, i.e. $\hat{U}_n^{(l)} = \widehat{Var}( \hat{U}_n^{(l)} ) $ for $l = 1, \dots, m$, a combination of them is given by $\bar{Q}_m := 1/m \sum_{l = 1}^{m} \hat{Q}_n^{(l)}$ and $\bar{U}_m :=  1/m  \sum_{l = 1}^{m} \hat{U}_n^{(l)}$ \citep{rubin2004multiple}. \cite{rubin2004multiple} suggested to use as the total variance of $(\bar{Q}_m - \mathbf{c}^\top \boldsymbol{\mu})$ the estimate $T_m = \bar{U}_m  + (1 + 1/m)B_m$, where $B_m$ is the variance between the $m$ complete-data estimates, i.e. $B_m = (m-1)^{-1} \sum_{l = 1}^{m} ( \hat{Q}_n^{(l)}  - \bar{Q}_m)^2$. The combined test statistic for the matched pairs design can then be approximated by a t reference distribution with $\nu_m$ degrees of freedom, i.e. 
\begin{align}
\label{Rubin}
\frac{\bar{Q}_m - \mathbf{c}^\top \boldsymbol{\mu}}{\sqrt{T_m} }  \quad \stackrel{H_0}{\approx} \quad  t_{\nu_m}, 
\end{align}
where $\nu_m = (m-1) (1+ r_m^{-1})^2$ and $r_m = (1+ m^{-1})B_m/\bar{U}_m$ being the relative increase in variance due to missingness. However, under small samples and a modest proportion of missing data, the approximation of the degrees of freedom for the t reference distribution in $(\ref{Rubin})$ might be inappropriate \citep{SmallRubin}. Adjusted degrees of freedom for $(\ref{Rubin})$ are given by 
\begin{align}
\label{RubinAdjusted}
\tilde{\nu}_m =  \nu_m \left( 1 + \frac{\nu_m}{ \hat{\nu}_{obs } }\right)^{-1},
\end{align}

where $\hat{\nu}_{obs} = n(n-1) [ (1 + r_m)(n+2) ]^{-1} $. In the sequel, we will restrict our attention to the choice given in $(\ref{RubinAdjusted})$, since $\tilde{\nu}_m$ is always smaller than $\nu_m$ and coincides with $\nu_m$, if the sample size is sufficiently large \citep{SmallRubin}. This is especially the case when considering datasets with sample size vectors of the form $(n_1, n_2, n_3) = k\cdot (30,10,10)$ for larger $k \in \{1, 2, \dots, 50 \}$ in Section \ref{sec5}. So far, theoretical guarantees for the validity of (\ref{PairedApprox}) resp. (\ref{RubinAdjusted}) using the \texttt{missForest} imputation procedure have not been delivered. Hence, it is completely unclear whether the paired t-test under the random forest based imputation  scheme will lead to valid inference.

	\section{Permuting Incomplete Data}
\label{sec4}
\cite{maritz1995permutation} and \cite{yu2012} proposed permutation techniques for incomplete paired data. However, these methods have the drawback that certain distributional assumptions (such as $0$-symmetry) are required to develop a valid level $\alpha$ test even at least asymptotically. A novel permutation method overcame these problems while preserving the finite sample exactness property under similar assumptions \citep{amro2017permuting, amro2018multiplication}. Besides, the suggested method has the advantage that it is asymptotically robust against heteroscedasticity and skewed distributions. The idea of \cite{amro2017permuting, amro2018multiplication} is based on permuting the complete and the incomplete cases separately and then combining the results using either a general weighted test statistic or a multiplication combination test involving both, the paired t-test and the Welch-type test statistic.\\

Based on a simulation study \citep{amro2018multiplication}, the two permutation methods showed similar behavior in most of the considered cases. Therefore, we only consider the method of \cite{amro2017permuting} for further analysis. Their suggested test statistic consists of the following weighting scheme:
\bqan \label{T}
T_{ML} = \sqrt{a} T_{t}(\mathbf{X}^{(c)}) + \sqrt{1-a} T_{w}(\mathbf{X}^{(i)})
\eqan
where, $a\in [0,1]$ is used to weight the inference from the complete pairs and the incomplete ones. Moreover, $T_{t}$ is the usual paired t-type statistic, $ T_{t}= T_{t}(\mathbf{X}^{(c)}) = (\sum_{i=1}^{n_1}  \mathbf{c}^\top \mathbf{X}_i^{(c)} )/(\sqrt{  n_1}\whsigma_1 )$,  $\mathbf{c}^\top \mathbf{X}_i^{(c)}=X_{1i}^{(c)}-X_{2i}^{(c)}$ denote the differences of the first and second component for $i=1,\ldots,n_1$ while 
$\whsigma^2_1=(n_1-1)^{-1}\sum_{i=1}^{n_1} (\mathbf{c}^\top \mathbf{X}_i^{(c)} -\overline{\mathbf{c}^\top \mathbf{X}_\cdot^{(c)} })^2$ is their empirical variance. Furthermore, $T_{w}$ is the Welch-type test statistic      
$T_{w}= T_{w}(\mathbf{X}^{(i)}) = (\overline{X}_{1\cdot}^{(i)} - \overline{X}_{2\cdot}^{(i)})/(\sqrt{  \frac{\whsigma_2^2}{n_2} + \frac{ \whsigma_3^2}{n_3} }),$	where $\overline{X}_{1\cdot}^{(i)} = n_2^{-1}\sum_{j=1}^{n_2} X_{1j}^{(i)}$ and $\overline{X}_{2\cdot}^{(i)} = n_3^{-1}\sum_{j=1}^{n_3} X_{2j}^{(i)}$ are the sample means and 
$\whsigma_2^2=  (n_2-1)^{-1}\sum_{j=1}^{n_2} (X_{1j}^{(i)} - \overline{X}_{1\cdot}^{(i)})^2$ and 
$\whsigma_3^2 =  (n_3-1)^{-1}\sum_{j=1}^{n_3} (X_{2j}^{(i)} - \overline{X}_{2\cdot}^{(i)})^2$ the corresponding empirical variances. A recommended value of  $a$ is $2n_1/(n+n_1)$ \citep{amro2017permuting}. This choice guarantees the correct treatment in the extreme scenarios of $n_1=n$ and $n_1=0$. In order to derive empirical quantiles of the test statistic $T_{ML}$, a separate permutation technique for the complete $\mathbf{X}^{(c)}$ and incomplete pairs $\mathbf{X}^{(i)}$ has been used. While the components of the complete cases are permuted independently, for the missing parts $\mathbf{X}^{(i)}$, the components are shuffled together \citep{amro2017permuting}. This method has the advantage of using all observed information while not being based on any parametric assumption. In particular, it was shown to be asymptotically correct in general and even finitely exact, if the distribution of the data is invariant with respect to the considered permutations (keyword: exchangeable in a certain way).

	\section{Simulation strategy}
\label{sec5}
In our simulation study, we compare the methods described in Section \ref{sec3} - \ref{sec4} with respect to their 
\begin{enumerate}[(i)]
	\item type-I-error rate control at level $\alpha = 5\%$ and 
	\item  power to detect deviation from the null hypothesis. 
\end{enumerate}

We thereby restrict the analysis to the paired t-test based on four imputation techniques; The established and easy to use MICE approach by \cite{buuren2010mice} with NORM and PMM as well as the \texttt{missForest} procedures 
RF MI and RF MICE described in Section \ref{sec3}. The latter was shown to be the method of choice in terms of accuracy in several simulation studies \citep{stekhoven2011missforest, waljee2013comparison, ramosaj2017wins}. The crucial simulation study was conducted within the statistical computing environment $\textsf{R}$ based on $n_{sim} = 10,000$ Monte-Carlo runs. Small to large sized paired data samples are generated by
\begin{center} 
	$\mathbf{X}_j = \boldsymbol{\Sigma}^{\frac{1}{2}} \boldsymbol{\epsilon}_j + \boldsymbol{\mu},\quad  j=1,... ,n ,$
\end{center}
where $\boldsymbol{\epsilon}_j = [\epsilon_{1j}, \epsilon_{2j}]^\top$ is an i.i.d. bivariate random vector with zero mean and identity covariance matrix.
Different choices of symmetric as well as skewed residuals are considered such as standardized normal, exponential, Assymetric Laplace or the $\chi^2$ distribution with $df = 30$ degrees of freedom. For the covariance matrix $\boldsymbol{\Sigma}$, we 
considered a homoscedastic and a heteroscedastic setting given by the choices  
\begin{center}
	$ \boldsymbol{\Sigma_1} =  \begin{bmatrix}
	1 & \rho \\
	\rho & 1
	\end{bmatrix} \text{ and } \boldsymbol{\Sigma_2} =  \begin{bmatrix}
	1 & 2\rho \\
	2\rho & 4
	\end{bmatrix} \textbf{},$
\end{center} 
for varying correlation factors $\rho \in (-1,1)$. Missing values are generated under the MCAR scheme by randomly inserting them to  the first or second component of the bivariate vector $\mathbf{X}_{j}$ until a fixed amount of missing values of size $n_2$ for the second and $n_3$  for the first component is achieved. Furthermore, the sample sizes were chosen as $(n_1,n_2,n_3)\in \{(10,10,10),(30,10,10)\}$ and later increased by adding $k \mathbf{1}_3^\top$ for $k\in \{1,...,500\}$ to the balanced sample size $(10,10,10)$ and multiplying the unbalanced setting $(30,10,10)$ with a factor $k\in \{1,...,50\}$. In order to assess the power of the paired-t-test for imputation methods and the  permutation test of \cite{amro2017permuting}, we set $\boldsymbol{\mu} = [\delta, 0]^\top$ with shifting parameter $\delta \in \{ 0, 1/2, 1 \}$. Here,  the permutation method was based on  $B=1,000$ permutation runs. Furthermore, we used $m = 5$ imputations  for the multiple imputation settings. 
In addition to reporting type-I-error ($\delta=0$) and power of the tests, we also report the NRMSE of the imputation methods defined by

\begin{equation*}
\begin{split}
\begin{gathered}
NRMSE =\\ \sqrt{\frac{ \sum\limits_{l = 1}^m \sum\limits_{j \in I_{n_2} } (X_{1j}^{true} - X_{1j}^{imp(l)} )^{2} +  \sum\limits_{j \in I_{n_3} } (X_{2j}^{true} - X_{2j}^{imp(l)} )^{2} }{ m  \left(\sum\limits_{j \in I_{n_2} } (X_{1j}^{true} - \bar{X}_{1\cdot}^{true} )^{2} +  \sum\limits_{j \in I_{n_3} } (X_{2j}^{true} - \bar{X}_{2\cdot}^{true} )^{2} \right)}}
\end{gathered}
\end{split}
\end{equation*}

Where, $X_{ij}^{true}$, is the true value whenever missing values are present, $\bar{X}_{i\cdot}^{true} = \frac{1}{\lvert I_{n_{i+1}} \rvert} \sum\limits_{j \in I_{n_{i+1}}} X_{ij}^{true}$, and $X_{ij}^{imp(l)}$ with $j \in I_{n_{i+1}}$, $i = 1,2$ is the imputed value of missing observations in the $l$-th iteration using a multiple imputation procedure \citep{rubin2004multiple}. 

\begin{figure*}[!t]
	\centering
	\includegraphics[scale = 1.2]{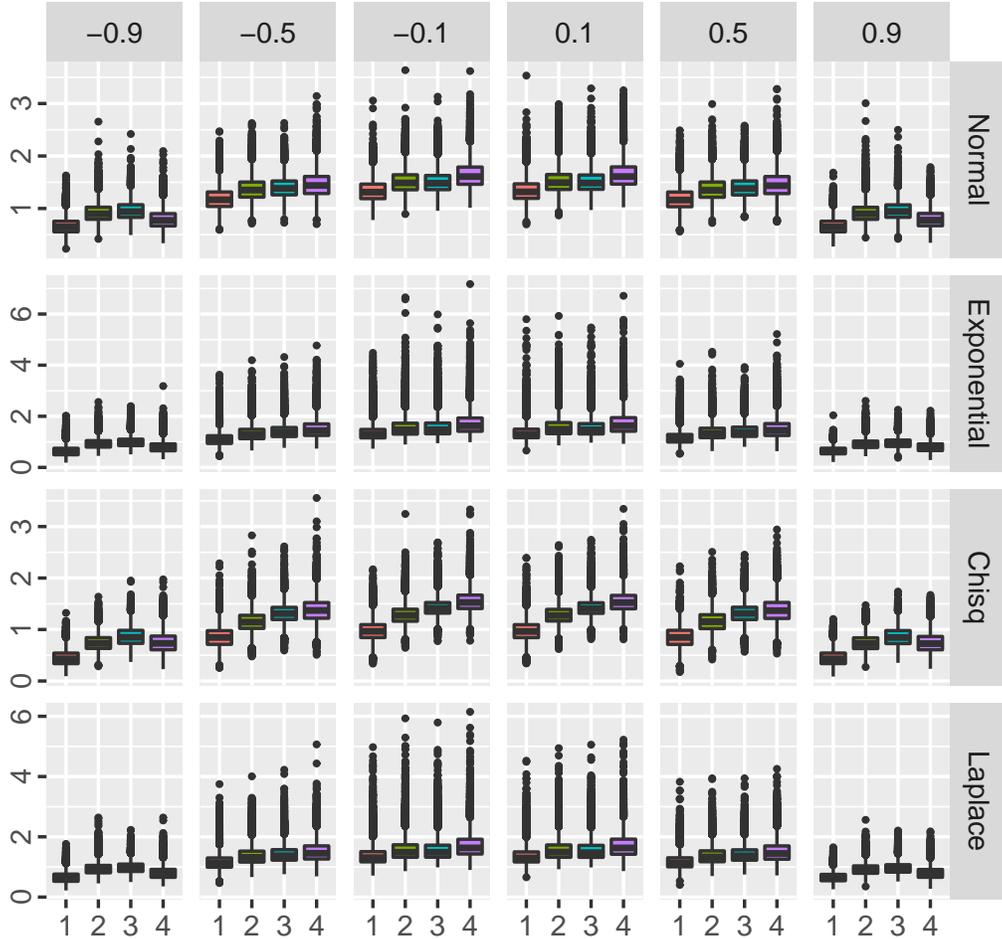}
	\caption{ NRMSE for the Normal, Exponential, $\chi_{df = 30}^2$ and Laplace distribution using $10,000$ Monte-Carlo runs under $H_0$ and the balanced scheme with $n_1 = n_2 = n_3 = 10$ under various correlations $\rho \in \{-0.9, -0.5, -0.1, 0.1, 0.5, 0.9 \}$ for (1) RFMI, (2) RFMICE, (3) PMM and (4) NORM.  }
	\label{NRMSE}
\end{figure*}

\begin{figure*}
	
	\begin{subfigure}{.5\textwidth}
		\centering
		\includegraphics[width=.8\linewidth]{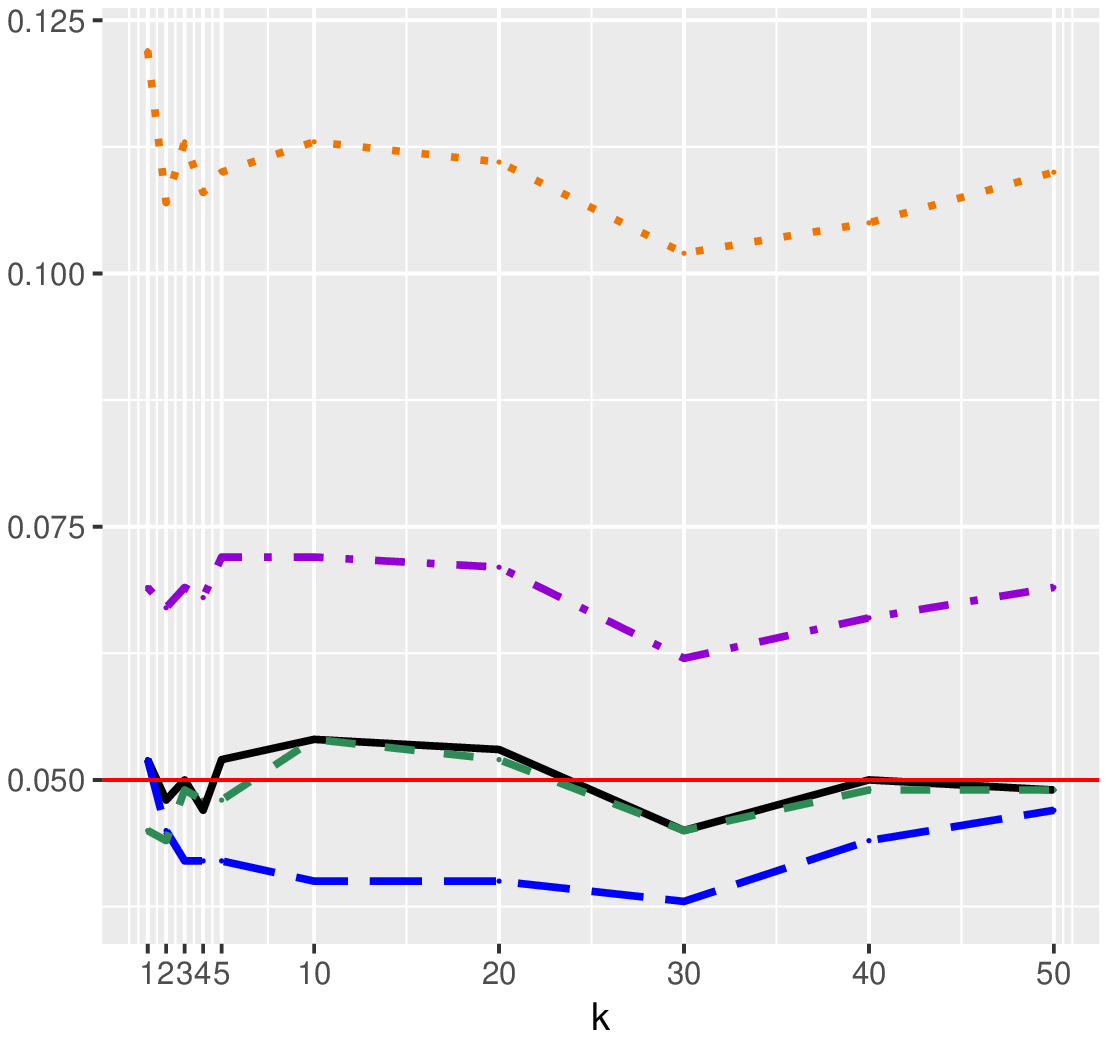}
		\caption{$k \cdot (30,10,10)$}
		\label{fig:sfig1}
	\end{subfigure}%
	\begin{subfigure}{.5\textwidth}
		\centering
		\includegraphics[width=.8\linewidth]{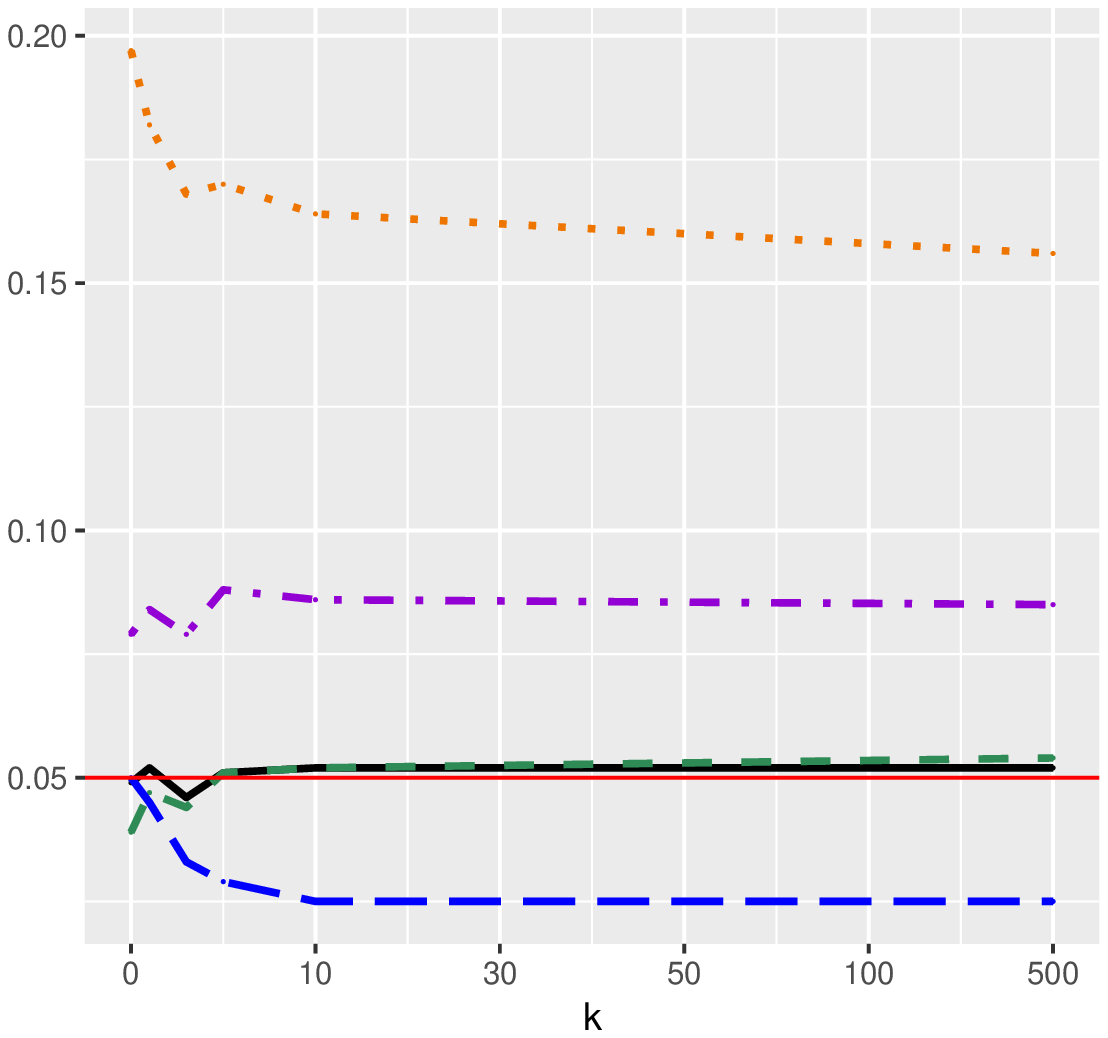}
		\caption{$k+(10,10,10)$}
		\label{fig:sfig2}
	\end{subfigure}
	\caption{Simulation results for type-I error level ($\alpha=0.05$) for  $T_{ML}$ (\textbf{\textendash\textendash\textendash}), $RF MI$({\color{bronze} \textbf{$\cdots$}}), $RF MICE$({\color{byzantine} \textbf{$-\cdot-$}}), $PMM$({\color{blue}\textbf{-- --}}),  $NORM$ ({\color{ao(english)} \textbf{- - -}}) for $\chi^2$ distribution under $\rho=0.1$ for varying $k$ values either multiplied to $(n_1,n_2,n_3)=(30,10,10)$ (a) or added to $(n_1,n_2,n_3)=(10,10,10)$ (b).}
	\label{fig:nk}
\end{figure*}

\begin{table*}[!t]
	\centering
	\caption{Simulation results for type-I error level ($\alpha=0.05$) for different distributions under varying correlation values $\rho$ each with sample sizes $(n_1,n_2,n_3)=(10,10,10)$ and different covariance matrices $\Sigma_1$ (homoscedastic) and $\Sigma_2$ (heteroscedastic).
		\label{Table1}}
	{\tabcolsep=3pt\begin{tabular*}{\textwidth}{@{\extracolsep{\fill}}lcccccccc c ccccc}
			Dist & $\rho$ & \multicolumn{5}{c}{$\Sigma_1$}  && \multicolumn{5}{c}{$\Sigma_2$} \\ \cline{3-7} \cline{9-13}
			&& $T_{ML}$ & RF MI & RF MICE & PMM & NORM
			&& $T_{ML}$ & RF MI & RF MICE & PMM & NORM \\
			
			Normal	&-0.9&0.052&0.070&0.058&0.061&0.045&&0.056&0.073&0.063&0.062&0.047\\
			&-0.5&0.052&0.118&0.071&0.061&0.045&&0.054&0.117&0.073&0.062&0.045\\
			&-0.1&0.050&0.163&0.077&0.055&0.038&&0.054&0.162&0.077&0.057&0.042\\
			&0.1&0.054&0.201&0.085&0.055&0.041&&0.064&0.188&0.085&0.051&0.041\\
			&0.5&0.052&0.260&0.085&0.048&0.039&&0.052&0.216&0.078&0.047&0.040\\
			&0.9&0.051&0.302&0.048&0.036&0.038&&0.058&0.167&0.054&0.046&0.042\\
			\\
			Exp	&-0.9&0.049&0.070&0.060&0.059&0.046&&0.074&0.085&0.075&0.076&0.058\\
			&-0.5&0.052&0.118&0.071&0.058&0.045&&0.074&0.124&0.078&0.065&0.051\\
			&-0.1&0.048&0.165&0.076&0.052&0.039&&0.077&0.173&0.088&0.063&0.045\\
			&0.1&0.052&0.195&0.078&0.050&0.040&&0.083&0.190&0.091&0.065&0.052\\
			&0.5&0.050&0.262&0.084&0.048&0.040&&0.084&0.221&0.092&0.06&0.056\\
			&0.9&0.045&0.295&0.043&0.034&0.038&&0.097&0.183&0.065&0.056&0.049\\
			\\
			Chisq	&-0.9&0.049&0.070&0.060&0.059&0.046&&0.054&0.073&0.062&0.061&0.046\\
			&-0.5&0.052&0.118&0.071&0.058&0.045&&0.055&0.115&0.073&0.056&0.041\\
			&-0.1&0.048&0.165&0.076&0.052&0.039&&0.052&0.167&0.079&0.055&0.044\\
			&0.1&0.052&0.195&0.078&0.050&0.040&&0.054&0.187&0.081&0.054&0.042\\
			&0.5&0.050&0.262&0.084&0.048&0.040&&0.058&0.219&0.082&0.057&0.049\\
			&0.9&0.045&0.295&0.043&0.034&0.038&&0.062&0.166&0.053&0.046&0.039\\
			\\
			Ass. &-0.9&0.050&0.070&0.060&0.061&0.045&&0.065&0.076&0.066&0.066&0.075\\
			Laplace&-0.5&0.051&0.108&0.067&0.054&0.040&&0.065&0.118&0.077&0.062&0.047\\
			&-0.1&0.050&0.167&0.077&0.055&0.040&&0.069&0.170&0.083&0.061&0.048\\
			&0.1&0.056&0.194&0.083&0.050&0.042&&0.069&0.191&0.083&0.057&0.046\\
			&0.5&0.052&0.252&0.082&0.046&0.046&&0.070&0.217&0.085&0.053&0.044\\
			&0.9&0.049&0.278&0.037&0.031&0.047&&0.090&0.182&0.061&0.051&0.049\\
			\bottomrule
		\end{tabular*}}{}
	\end{table*}

	\begin{table*}[!t]
		\centering
		\caption{Power simulation results ($\alpha=0.05$) for different distributions under varying correlation values $\rho$ for $(n_1,n_2,n_3)=(10,10,10)$ and for $\Sigma=\Sigma_1$ (homoscedastic).
			\label{Table2}}
		{\tabcolsep=3pt\begin{tabular*}{\linewidth}{@{\extracolsep{\fill}}lcccccccc c ccccc}
				Dist & $\rho$ & \multicolumn{5}{c}{$\delta=0.5$}  && \multicolumn{5}{c}{$\delta=1$} \\ \cline{3-7} \cline{9-13}
				&& $T_{ML}$ & RF MI & RF MICE & PMM & NORM
				&& $T_{ML}$ & RF MI & RF MICE & PMM & NORM \\
				
				Normal	  &-0.9&0.260&0.315&0.299&0.300&0.264&&0.733&0.792&0.780&0.780&0.747\\
				&-0.5&0.271&0.396&0.319&0.284&0.239&&0.768&0.847&0.803&0.771&0.719\\
				&-0.1&0.306&0.477&0.349&0.290&0.238&&0.832&0.909&0.846&0.788&0.734\\
				&0.1&0.338&0.551&0.383&0.306&0.250&&0.858&0.934&0.869&0.799&0.729\\
				&0.5&0.439&0.703&0.463&0.345&0.302&&0.947&0.985&0.924&0.864&0.805\\
				&0.9&0.869&0.967&0.657&0.605&0.628&&1.000&1.000&0.973&0.967&0.957\\
				\\
				Exp				&-0.9&0.287&0.332&0.312&0.315&0.269&&0.777&0.814&0.800&0.801&0.760\\
				&-0.5&0.336&0.435&0.364&0.318&0.271&&0.818&0.862&0.825&0.785&0.742\\
				&-0.1&0.379&0.516&0.394&0.328&0.269&&0.863&0.909&0.857&0.796&0.745\\
				&0.1&0.408&0.570&0.414&0.332&0.279&&0.888&0.928&0.861&0.813&0.756\\
				&0.5&0.528&0.727&0.496&0.380&0.343&&0.952&0.978&0.918&0.841&0.810\\
				&0.9&0.876&0.950&0.676&0.617&0.661&&1.000&0.998&0.968&0.954&0.958\\
				\\
				Chisq		&-0.9&0.255&0.311&0.290&0.295&0.254&&0.739&0.796&0.783&0.784&0.752\\
				&-0.5&0.278&0.391&0.322&0.287&0.243&&0.776&0.848&0.810&0.774&0.729\\
				&-0.1&0.315&0.500&0.367&0.301&0.248&&0.831&0.905&0.843&0.786&0.729\\
				&0.1&0.345&0.550&0.390&0.308&0.256&&0.874&0.938&0.873&0.809&0.748\\
				&0.5&0.442&0.700&0.448&0.342&0.304&&0.951&0.985&0.927&0.856&0.802\\
				&0.9&0.868&0.966&0.661&0.610&0.636&&1.000&1.000&0.970&0.966&0.957\\
				\\
				Ass.&-0.9&0.287&0.336&0.315&0.316&0.270&&0.767&0.812&0.791&0.792&0.753\\
			Laplace	&-0.5&0.327&0.421&0.354&0.317&0.263&&0.814&0.859&0.822&0.791&0.739\\
				&-0.1&0.367&0.519&0.385&0.322&0.267&&0.858&0.908&0.851&0.798&0.743\\
				&0.1&0.400&0.572&0.412&0.336&0.277&&0.889&0.935&0.874&0.812&0.754\\
				&0.5&0.504&0.726&0.497&0.375&0.342&&0.952&0.979&0.920&0.846&0.815\\
				&0.9&0.877&0.950&0.669&0.614&0.660&&1.000&0.999&0.971&0.955&0.959\\
				\bottomrule
			\end{tabular*}}{}
		\end{table*}

The NRMSE has been reported as a key measure for assessing imputation quality of various imputation techniques \citep{stekhoven2011missforest, waljee2013comparison,ramosaj2017wins}. However, it is completely unclear whether an imputation technique with relatively low NRMSE automatically lead to type-I-error rate control or higher power in incompletely observed matched pairs designs.  Below, we shed light into this obstacle.

	\subsection{NRMSE Results}
	\label{sec5_1}
	The simulation was conducted under $H_0$ with homogeneous covariance matrix and a sample size of $n_1 = n_2 = n_3 = 10$ for different correlation coefficients and different distributions as described above. It can be readily seen from Figure~\ref{NRMSE} that RF MI resulted in the lowest NRMSE overall settings under consideration. For low to moderate correlation factors (up to $|\rho|=0.5$) RF MICE showed the second best overall NRMSE, followed by PMM and NORM. For larger correlation factors $|\rho|=0.9$ this ordering changes with NORM showing the second lowest NRMSE, followed by RF MICE and NORM.
	Anyhow, our findings are in line with previous results \citep{waljee2013comparison, ramosaj2017wins} in that they also result in a recommendation to apply RF MI if NRMSE is the major assessment criterion. However, we will see below that
	NRMSE is not a suitable measure for evaluating an imputation method when aiming to conduct inference on small to moderate sample sizes; at least in matched pairs designs.
	\subsection{Type-I- Error Control Results}
	\label{sec5_2}
	Simulation results of type-I-error of the $T_{ML}$- permutation test as well as the paired t-test based on imputation are summarized in Table \ref{Table1} under various covariance structures and residual distributions for $(n_1,n_2,n_3)=(10,10,10)$. Starting with the permutation test $T_{ML}$ of \cite{amro2017permuting} we see that it tends to be an adequate exact level $\alpha$ test among all considered ranges of $\rho$ for homoscedastic covariance structures and most heteroscedastic settings. 
	Only in the heteroscedastic cases with skewed exponential (and to some extend also for asymmetric Laplace) distributed residuals, the permutation test has some difficulties in controlling type-I-error rate. In those cases, the MICE procedure under the algorithmic implementation of NORM has shown favorable type-I error rate control, but this with the cost of considerably less power (Table~\ref{Table2}). For other residual distributions, NORM obtains the significance level of $\alpha = 0.05$ well, while being slightly conservative, especially when the correlation structure turns positive. This effect of a decrease in type-I error with increasing positive correlation factors $\rho$ could also be obtained for the PMM procedure and partly for the RF MICE. Here, the permutation test $T_{ML} $ seems not to be affected by the various degrees of correlations. In contrast to the permutation test and the MICE procedures, both RF MI and RF MICE, imputation procedures based on Random Forest models, failed to control type-I error rate in almost all settings. Especially RF MI, which is based on repeatedly using \texttt{missForest}, heavily inflates type-I error of the paired t-test. RF MICE, which introduces more variability in the imputation draws, could be able to reduce type-I error, but is only in strongly positive correlated settings able to drop below the $0.05$ threshold. 

		In addition to these small sample size settings, we are also interested in type-I error rate control when sample sizes increases, while missing rates remain nearly unchanged. For moderate to large sample sizes, we considered the choices $k \cdot (30,10,10)$ and $k + (10,10,10)$, where $k$ ranges from $1 - 500$ (balanced case) and $1 - 50$ (unbalanced), respectively. Figure $\ref{fig:nk}$ summarizes type-I error rate control at $\alpha = 0.05$ for these settings.
		The results indicate that the imputation procedures based on Random Forest models, RF MI and RF MICE, failed to control type-I error rate also in larger samples. This especially holds for RF MI with type-I error rates between $11\%$ and $16\%$ for the largest sample size setting ($n>1,500$) while RF MICE was only slightly liberal then ($6.5 - 8\%$). 
		Only the permutation test $T_{ML}$ and the MICE procedure under the algorithmic implementation of NORM control the type-I error accurate for moderate to large sample sizes, while PMM even got more conservative. 
		\begin{figure*}
			\centering
			\fbox{\includegraphics[width=0.45\textwidth,scale=0.7]{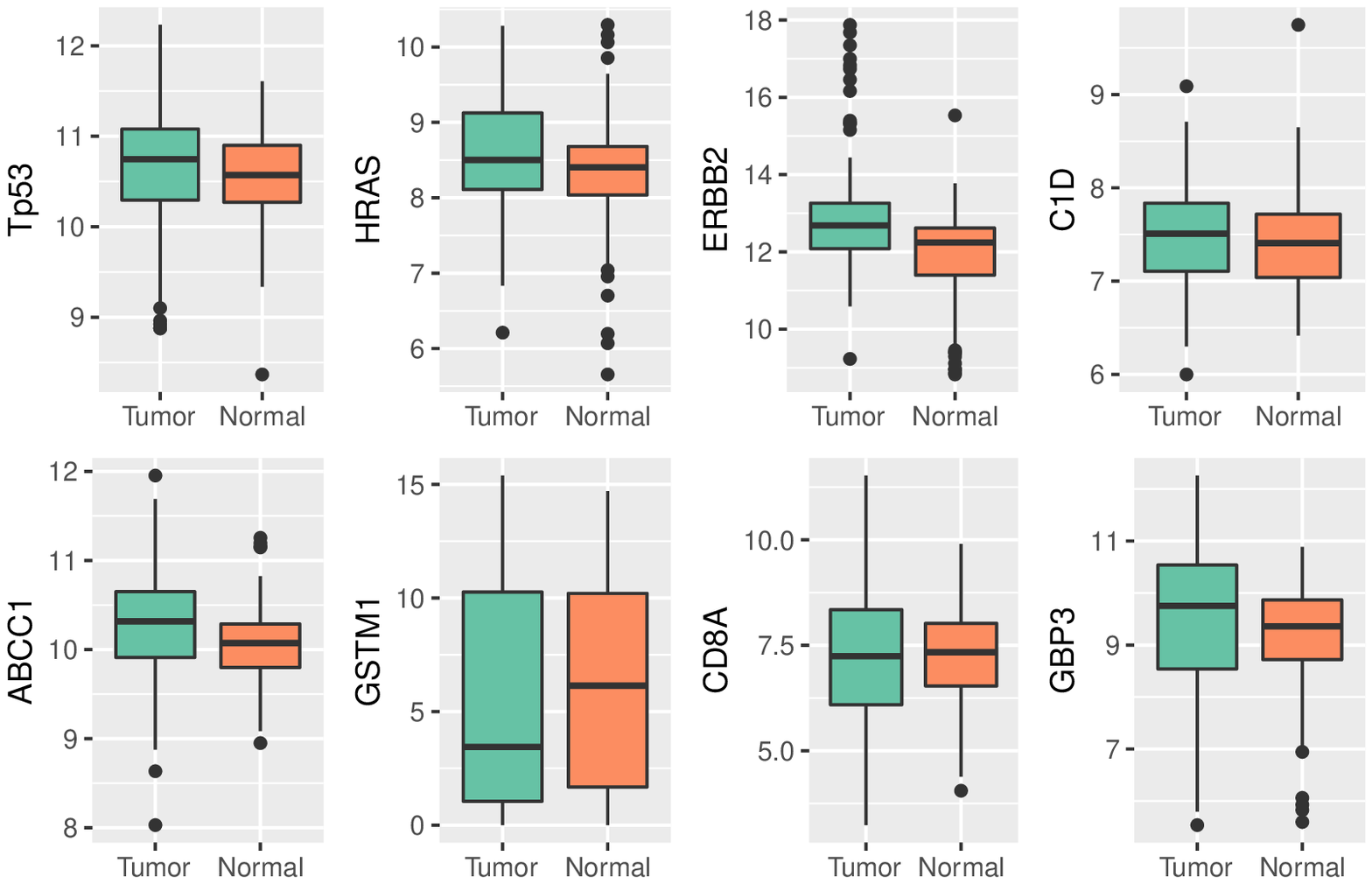}}
			\caption{Box plots of the Gene expression levels of the tumor and normal breast tissues for the eight different genes TP53, ABCC1, HRAS, GSTM1, ERBB2, CD8A, C1D and GBP3. \label{genes} }
		\end{figure*}	
		\subsection{Power Results}
		\label{sec5_3}
		In addition to the type-I-error rate, we simulated the power of the permutation test and the paired t-test based on the four imputation approaches. Table \ref{Table2} summarizes the power analysis for the sample sizes $(n_1,n_2,n_3)=(10,10,10)$ under the range of $\rho$ and for a homoscedastic covariance structure. Power simulations for the other combinations of sample sizes and covariance structures were similar, see the supplement for details. It can be seen from Table \ref{Table2} that RF MI combined with Rubin's Rule has the largest power among all considered tests and procedures. However, this is not remarkable due to its extreme liberal behavior. Thereafter, RF MICE and the permutation test $T_{ML}$ performed comparable in terms of power.
		Only in case of large positive correlation $T_{ML}$ performed considerably better. Contrary, parametric imputation methods such as PMM and NORM resulted into much lower statistical power. For example NORM, which can be seen as very competitive to the permutation test in terms of type-I error rate control, often exhibits $10\%-20\%$ less power than $T_{ML}$. 
		Hence, the permutation test qualifies as a suitable test statistic throughout different correlations while controlling type-I-error and delivering a comparably high power in matched pairs designs with partially observed data.

	\section{Analysis of the Cancer Genome Atlas Dataset}
\label{sec6}

We consider data from a breast cancer study to illustrate the behavior of the considered imputation methods as opposed to the permutation method in a clinical study. Our data comes from the Cancer Genome Atlas (TCGA) project, which was launched in 2005 with a financial support from the National Institutes of Health. It aims to understand the genetic basis of several types of human cancers through the application of  high-throughput genome analysis techniques. TCGA is willing to improve the ability of diagnosing, treating and even preventing cancer through discovering the genetic basis of carcinoma. Molecular information such as mRNA/miRNA expressions, protein expressions, weight of the sample as well as clinical information about the  participants were collected. Clinical and RNA sequencing records could be obtained from $1093$ breast cancer patient, where $112$ of them provided both, normal and tumor tissues. We used the $\log_2$-transformed data set for gene expressions. Luckily, no missing values were present, so that the data could be analyzed by means of the classical paired t-test. The results are then compared with the above procedures after artificially inserting missing values as follows: we generated Bernoulli distributed random variables $R_{ij}\sim Bernoulli(r)$ such that $\mathbf{X}_j^{(o)} = [R_{1j} \cdot X_{1j}, R_{2j} \cdot  X_{2j}]^\top$ with $r = 0.2$ leading to an overall missing rate of $30\%$.\\

\begin{table}[!t]
	\caption{Description of the genes that were selected as auxiliary variables.\label{AUXtable}}%
	\begin{tabular}{|L{1.5cm}|L{4cm}| L{9cm}|}
			\hline
			\textbf{Gene}&\textbf{Group$^*$}&\textbf{Selected Auxiliary Gene Expressions} \\
			\hline 
			\hline
			\textbf{TP53}&Tumor Protein p53
			& TPCN2, TPD52L2, TPST1, TP63, TP53TG1, TPD52L1, TPRKB, TP53INP2, TP53BP1, TPRN, TPM4, TPRA1, TPP2, TPPP3, TPSB2, TPT1, TPM3, TPO, TPK1, TP53INP1, TPBG, TPX2, TPTE2P1, TPSAB1, TPMT, TP53BP2, TPR, TPRG1, TP53I13, TPCN1, TPI1P2, TPI1, TPRA1, TPM2, TPM1, TPST2, TPR, TPI1, TP53INP2, TPRN\\\hline
			\textbf{ABCC1}&Multidrug Resistance-associated Protein 1  &ABCC10, ABCC11, ABCC3, ABCC4, ABCC5, ABCC6P2, ABCC6, ABCC9, ABCC10, ABCC11, ABCC2,  ABCC3, ABCC4, ABCC5, ABCC6P2, ABCC6, ABCC9\\\hline
			\textbf{HRAS}&Transforming Protein p21 &HRASLS5, HRASLS2, HRASLS5, HRASLS\\\hline
			\textbf{GSTM1}&Glutathione S-transferase Mu 1 &GSTA4, GSTCD, GSTK1, GSTM2, GSTM3, GSTM4, GSTM5, GSTO1, GSTO2, GSTP1, GSTT2, GSTZ1, GSTA4, GSTCD, GSTK1, GSTM2, GSTM3, GSTM4, GSTM5, GSTO1, GSTO2, GSTP1, GSTT2, GSTZ1\\\hline
			\textbf{ERBB2}&Erb-B2 Receptor Tyrosine Kinase 2&ERBB2IP, ERBB3, ERBB4, ERBB2IP, ERBB3, ERBB4\\\hline
			\textbf{CD8A}&CD8a Molecule&CD80, CD81, CD82, CD83, CD84, CD86, CD8B, CD80, CD81, CD82, CD83, CD84, CD86, CD8B\\\hline
			\textbf{C1D}&Nuclear Nucleic Acid-Binding Protein&PRKD1, PRKD2, PRKD3, PRKDC, TSNAX, PRKD1, PRKD2, PRKD3, PRKDC, TSNAXIP1, TSNAX\\\hline
			\textbf{GBP3}&Guanylate-Binding Protein &GBP1, GBP2, GBP4, GBP5, GBP1, GBP2, GBP4, GBP5\\\hline
			\bottomrule
		\end{tabular}
		*The sceintific names of the genes were provided from GeneCards; http://www.genecards.org.\footnotemark\\
		\footnotetext{Footnote}
	\end{table}
	Based on previous studies, six genes have been identified to be significantly related to breast cancer: \textbf{TP53, ABCC1, HRAS, GSTM1, ERBB2} and \textbf{CD8A} \citep{finak2008stromal, harari2000molecular, munoz2007role, de2002genes}. Another two genes; \textbf{C1D} and \textbf{GBP3} were under investigation but did not show any significant relation towards breast cancer patients \citep{qi2018testing}. In this paper, we aim to test the hypothesis whether mean genetic expressions of the eight genes differ significantly between normal and tumor tissues of breast cancer patients. Boxplots representing the characteristics of the genes are shown in Figure \ref{genes}. For more details regarding the dataset, we refer to Firehouse (www.gdac.broadinstitute.org).\\
	\begin{table}[!t]%
		\caption{Two-sided P-values of  the breast cancer study  restricted to only the considered genes.\label{pvalue}}%
		{\tabcolsep=11pt\begin{tabular*}{\linewidth}{@{}lllllll@{}}\toprule
				\textbf{Gene}&\textbf{T(Comp)}  & $\boldsymbol{T}_{\mathbf{ML} }$ &\textbf{RF MI} &\textbf{RF MICE}&\textbf{PMM} & \textbf{NORM}  \\\midrule
				\textbf{TP53}&0.153&0.054&\textbf{0.011}&\textbf{0.046}&0.056&\textbf{0.041} \\
				\textbf{ABCC1}&0.001&0.016&0.008&0.022&\textbf{0.085}&0.017\\
				\textbf{HRAS}&0.008&0.002&$2.4\cdot 10^{-5} $ &0.001&0.001&0.001\\
				\textbf{GSTM1}&0.002&0.008& $3.9\cdot 10^{-5}$&0.001&0.001&0.003\\
				\textbf{ERBB2}&$9.7\cdot 10^{-9}$&0.000&$9.3\cdot 10^{-11}$&$8.6\cdot 10^{-8}$&$3.6\cdot 10^{-6}$&$4.4\cdot 10^{-6}$\\
				\textbf{CD8A}&0.933&0.558&0.205&0.280&0.418&0.378\\
				\textbf{C1D}&0.414&0.480&0.278&0.287&0.489&0.529\\
				\textbf{GBP3}&0.006&0.01&\textbf{0.064}&\textbf{0.135}&\textbf{0.103}&\textbf{0.059}\\
				\hline
			\end{tabular*}}
			The bold figures represent false decisions (rejecting a true $H_0$ or failing to reject a false $H_0$ at $\alpha=0.05$)\footnotemark\\
			\footnotetext{Footnote} 
		\end{table}
		
		\begin{table}[!]
			\caption{Two-sided P-values of  the breast cancer study with auxiliary variables.\label{pvalueAUX}}%
			{\tabcolsep=16pt\begin{tabular*}{\linewidth}{@{}lllll@{}}\toprule
					\textbf{Gene}&\textbf{RF MI} &\textbf{RF MICE}&\textbf{PMM} & \textbf{NORM}  \\\midrule
					\textbf{TP53}&\textbf{0.021}&0.054&\textbf{0.04}&0.069\\
					\textbf{ABCC1}&0.001&0.009&0.011&0.007\\
					\textbf{HRAS}&$1.4\cdot 10^{-5}$&0.002&0.003&0.003\\
					\textbf{GSTM1}&0.015&\textbf{0.156}&0.001&0.011\\
					\textbf{ERBB2}&$4.1\cdot 10^{-11}$&$7.6\cdot 10^{-8}$&$2.5\cdot 10^{-7}$&$2.2\cdot 10^{-7}$\\
					\textbf{CD8A}&0.559&0.682&0.618&0.621\\
					\textbf{C1D}&0.217&0.292&0.297&0.364\\
					\textbf{GBP3}&0.006&\textbf{0.068}&0.021&\textbf{0.064}\\\hline
				\end{tabular*}}
				The bold figures represent false decisions (rejecting a true $H_0$ or failing to reject a false $H_0$ at $\alpha=0.05$).\footnotemark\\
				\footnotetext{Footnote}
			\end{table}
			The data set delivers in total $18,322$ gene expressions. From the viewpoint of an imputer, a rich data base that can be used to impute missing values is not only restricted to the paired observations of the gene expressions required for analysis. \cite{meng1994multiple}, for example, mentioned that additional information (called auxiliary variables in the sequel)  should be considered in an imputation method. Since in general, the number of available gene expression results into a $p > n$ problem, in which least square estimates used in PMM and NORM will suffer from instable estimation, restriction to a suitable subset is required. In Table \ref{AUXtable}, the gene expressions selected for auxiliary candidates are given. The selection of additional gene expressions was based on their membership to groups, that have been specified based on the biochemical structure of the eight genes. In addition, all gene expressions consisting of at least one missing values have been dropped, in oder to make the result comparable in the later analysis.  	
			
			All eight genes have been tested separately for mean equality between normal and tumor tissue using the 
			\begin{enumerate}[(i) ]
				\item  paired t-test in the complete-case analysis before artificially inserting missing values,
				\item  the paired t-test after (multiple) imputing missing values with RF MI, RF MICE, PMM and NORM,
				\item  the paired t-test after (multiple) imputing missing values with  RF MI, RF MICE, PMM and NORM using auxiliary variables, 
				\item  the weighted permutation test $T_{ML}$ after inserting missing values.
			\end{enumerate}
			
			The resulting $p$-values of the corresponding tests are summarized in Table \ref{pvalue} and \ref{pvalueAUX}.  In contrast to the simulation study, we extended the number of imputation draws to $m = 100$. The paired t-test in the complete-case analysis identified five out of eight genes having significantly different genetic expressions in normal and tumor tissues; genes 
			\textbf{ABCC1, HRAS, GSTM1, ERBB2,} and   \textbf{GBP3}. Mean gene expressions in \textbf{TP53, CD8A}, and \textbf{C1D}  have been identified as non significantly different between normal and tumor tissues. Same results for all eight genes could be obtained using the weighted permutation test $T_{ML}$. However, all considered imputation methods except PMM led to different results for the \textbf{TP53} gene indicating significantly different results in mean gene expressions between normal and tumor tissues. Even NORM, which indicated favorable results among the imputation procedures during simulation, failed to deliver similar results as the paired t-test in complete case analysis or $T_{ML}$. Here, including auxiliary variables for \textbf{TP53} gene helped NORM and also in case of RF MICE. However, this was not the case for RF MI or PMM. For example, the PMM test decisions were flipped
			after rerunning it with auxiliary variables, finding three more significant results of which one (\textbf{TP53}) was different to the the complete-case analysis before inserting missings. To discuss another example, all imputation techniques (RF MI, RF MICE, PMM and NORM) without auxiliary information had difficulties in rejecting the null hypothesis of equal mean gene expressions in \textbf{GBP3} gene leading again to different results compared to the paired t-test in complete-case analysis or $T_{ML}$. After including auxiliary variables, only the imputation tests RF MI and PMM found the same significance. A similar observation can be made for gene expression  \textbf{GSTM1},  where adding extra information lead to a non-rejection of the imputation test RF MICE.

	\section{Discussion}
		We investigated the effect of imputation procedures on inference in matched pairs designs. Our simulation results indicated that NRMSE is not a suitable measure for evaluating an imputation method with respect to subsequent inference. Although imputation methods based on the random forest have shown high predictive accuracy, they failed to control type-I-error under various settings and distributions. Moreover, even incorporating uncertainty in the test statistic by multiply imputing missing values as suggested by \cite{rubin2004multiple} failed to deliver better type-I-error control for the random forest based methods. Results turn into the right direction, when a modified version called RF MICE in the \texttt{mice} package is used. This because of the additional variance introduced in the random draws from observations falling in the corresponding leaf of a random decision tree. However, this imputation method realized a serious increase in NRMSE and a considerable loss of power while still exhibiting a rather liberal behavior under the null hypothesis. Parametric methods such as PMM and NORM in \texttt{mice} performed better in terms of type-I error rate control, but realized even less power to detect deviations from the null hypothesis. In contrast, a recently proposed permutation test that only works with the observed data at least delivered acceptable results and can be considered as the favorite choice when testing the equality of means in matched pairs with missing values.\\
		
		In a real data example with gene expressions form breast cancer patients, the effect of auxiliary variables in an imputation scheme has been considered as recommended by \cite{meng1994multiple}. For non-parametric imputation schemes based on random forest models, the effect of auxiliary variables were less effective compared to parametric choices such as PMM and NORM. Even-though, all the parametric and nonparametric imputation schemes; RF MI, RF MICE, PMM and NORM failed in mirroring all the significant and the non significant genes as obtained using the paired t-test in a complete case analysis before artificially inserting missing values. Contrary, the permutation procedure, capable of working with missings, led to the same conclusions. From these findings we recommend a cautionary use of imputation techniques (with or without auxiliary variables) for drawing inference. 	
		
		In summary, it is important to which further extent statistical analysis will be conducted when dealing with missing values. In terms of inference, imputation methods can fail to deliver suitable results as shown in the simulation and data example. Here, the permutation test of \cite{amro2017permuting} appeared to be the more reasonable choice.  However, if the statistical analysis on partially observed data aims to predict outcomes, imputation methods based on random forest models by \cite{stekhoven2011missforest} and \cite{ramosaj2017wins} could be a suitable solution. Future research will extend the usual investigation to more complex designs involving more than two endpoints and possibly multiple groups as well as the inclusion of more auxiliary variables. Regarding the latter we avoided a $p>n$ situation as parametric imputation methods such as PMM and NORM may run into difficulties. However, random forest models are known to be capable of dealing with high-dimensional feature problems and it would be interesting to also investigate this in case of inference after imputation.\\

	\section{Acknowledgments}
		We acknowledge the support of the Daimler AG represented by the research department 'Computer Vision and Camera Systems'. This  work  was also  supported  by  the  German  Academic  Exchange  Service  (DAAD)  under  the 	project: Research Grants - Doctoral Programmes in Germany, 2015/16 (No. 57129429). \\
		{\it Conflict of Interest}: None declared.

	\bibliographystyle{apalike}
	\bibliography{reference}{}

\end{document}